\documentclass[aip,rsi,11pt,reprint,superscriptaddress]{revtex4-1}

\usepackage{amsmath}    
\usepackage{graphicx}
\usepackage{hyperref}   

\usepackage{natbib}
\usepackage{natmove}

\renewcommand{\deg}{$^\circ$}

\newlength{\narrowf}
\newlength{\widef}
\begin{document}
	
	\title{X-ray tomography system to investigate granular materials during mechanical loading}
	
	\date{\today}

	\author{Athanasios~G.~Athanassiadis}
	\affiliation{James Franck Institute and Department of Physics, The University of Chicago}
	\affiliation{Department of Mechanical Engineering, Massachusetts Institute of Technology}   
    \author{Patrick J. La Rivi\`ere}
    \author{Emil Sidky}
    \affiliation{Department of Radiology, The University of Chicago}
    \author{Charles Pelizzari}
    \affiliation{Department of Radiation and Cellular Oncology, The University of Chicago}
    \author{Xiaochuan Pan}
    \affiliation{Department of Radiology, The University of Chicago}
	\author{Heinrich~M.~Jaeger}
	\email[Please send correspondence to: ]{h-jaeger@uchicago.edu}
	\affiliation{James Franck Institute and Department of Physics, The University of Chicago}
	\begin{abstract}

	We integrate a small and portable medical x-ray device with mechanical testing equipment to enable \emph{in-situ}, non-invasive measurements of a granular material's response to mechanical loading. We employ an orthopedic C-arm as the x-ray source and detector to image samples mounted in the materials tester. We discuss the design of a custom rotation stage, which allows for sample rotation and tomographic reconstruction under applied compressive stress. We then discuss the calibration of the system for 3D computed tomography, as well as the subsequent image reconstruction process. Using this system to reconstruct packings of 3D-printed particles, we resolve packing features with 0.52~mm resolution in a (60~mm)$^3$ field of view. By analyzing the performance bounds of the system, we demonstrate that the reconstructions exhibit only moderate noise.

	\end{abstract}

	\maketitle

	\section{Introduction}

	Typically, granular materials research is interested in the complex relationship between a packing's bulk behavior and its local, microstructural evolution. Commonly, this requires a simultaneous understanding of the forces acting on a granular system and the microstructural rearrangements that respond to these forces \cite{Duran1999,Andreotti2013,Jaeger1996,Herrmann1998}. However, studying the local evolution in three dimensions remains a challenge because most granular materials are optically opaque. With the exception of index-matched particles packed in a fluid \cite{Liu1995,Brujic2003,Dijksman2012}, \emph{in-situ} observation of local particle configurations in a three-dimensional (3D) packing requires specialized non-invasive imaging techniques such as magnetic resonance or x-ray imaging \cite{Nakagawa1993,Kuperman1996,Hill1997,Nakagawa1997,Oda1998,Seymour2000,Mueth2000,Seidler2000,Mobius2004,Aste2006,Scheel2008,Jerkins2008,Sanfratello2009,Zou2009,Jaoshvili2010,Delaney2010,Higo2011,Shepherd2012,Fu2012,Brown2012,Neudecker2013}. X-ray imaging, in particular, has proven to be one of the most versatile techniques to peer inside of the microstructure in 3D systems. To track time-dependent processes in a system's interior, 2D projection radiography has the advantage of speed. \cite{Baxter1989,Royer2005,Royer2007,Maladen2009,Royer2011,Waitukaitis2012}. However, to capture the full 3D structure of granular systems, x-ray tomography is required \cite{Aste2006,Zou2009,Jaoshvili2010,Delaney2010,Higo2011,Shepherd2012,Fu2012,Brown2012,Neudecker2013}.

	\begin{figure*}[t]
	\centering
	\includegraphics[width=.9\widef]{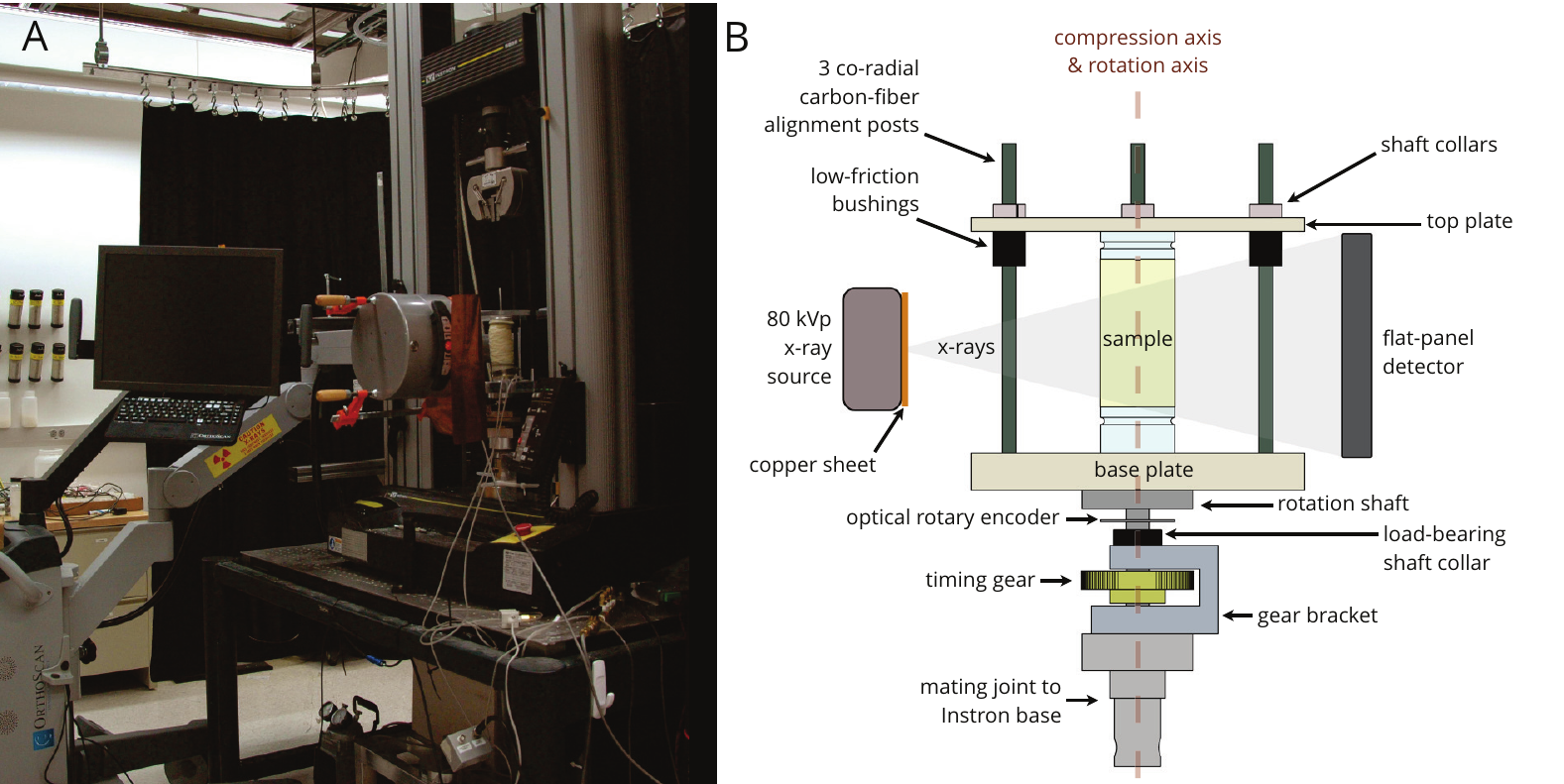}
	\caption{\label{fig:setup1} (A) Imaging setup located in the Instron materials tester, allowing for \emph{in-situ} measurements of microstructure and bulk response. (B) Detailed schematic of tomography rotation stage. }
	\end{figure*}

	When imaging with x-rays, there are trade-offs to using different imaging systems. Synchrotron sources can produce very high resolution images, and their brightness allows for imaging at high speed. However, the beam cross section is typically very small, constraining the imaging field-of-view \cite{Royer2005,Zou2009,Brown2012}. Medical x-ray scanners can provide a large field-of-view with high resolution. However, such systems are expensive and severely limit access to internal hardware and software configurations, making them difficult to integrate into experiments and extract optimized quantitative measurements of non-biological samples. Benchtop or larger CT scanners for materials science, can produce high resolution 3D reconstructions \cite{Higo2011,Shepherd2012,Neudecker2013}. However, these scanners are often limited to small samples (typically no larger than a couple centimeters tall) and usually require special, dedicated test fixtures to hold samples and apply stress.

	What has been lacking in granular mechanics research is a versatile x-ray system that can be combined with existing materials testing equipment \emph{and} can image the large samples used in such equipment (typically 50-75~mm in diameter and 100-150~mm tall). Here we describe such a combination, based on a small, mobile, medical x-ray system called a mini C-arm. In a C-arm, the x-ray source and detector are located at the two ends of a c-shaped metal beam that can be positioned as needed in 3D. These machines typically are used in image-guided medical procedures, where the open side of the ``C'' can easily be positioned to fit around the body part to be imaged. With large C-arms, computer controlled rotation of the ``C'' around the patient allows for 3D tomography \cite{Fahrig1997,Orth2008}; with mini C-arms, the beam position must be adjusted manually and tomography is not performed in a medical setting. Instead, mini C-arms are typically used to perform quick diagnostics or monitor the positioning of catheters during surgical procedures. Recently, medical C-arms have found use outside of hospitals as well; D. Goldman and colleagues used projection images from a mini C-arm to track the movement of small lizards ('sandfish') through granular media \cite{Maladen2009}. While these examples demonstrate how a fixed mini C-arm can be used for 2D-projection data, tomography of a static sample is also possible if the sample is rotated in front of the beam.

	Our x-ray tomography system integrates an Orthoscan mini C-arm with a custom sample rotation stage. It is combined with an Instron 5800 series materials tester that we use to strain the sample while measuring stress. This setup allows us to perform tomographic analysis of the local internal structure of samples at any stage of a mechanical loading process. Advantageously, this combination provides a large field of view - (6~cm)$^3$ - reconstructed to resolve 520~$\mu$m features, together with the ability to measure forces up to 50~kN and displacements as small as 10~$\mu$m. With this setup, we demonstrate the ability to track microstructural changes in relation to the exact stresses exerted on the sample.

	\section{Hardware}

	\subsection{Setup}

	To perform \emph{in-situ} compression and tomography, we designed a system around three key components: a materials tester (Instron 5869, Instron), a mini C-arm x-ray source and detector (Orthoscan FD, Orthoscan Inc.), and a custom rotation stage that we designed specifically for this system. The Instron materials tester sits on a lab bench and consists of a base on which to mount the sample, a horizontal cross-head with a load cell to compress the sample and measure the resulting forces, and two vertical pillars which control and measure the cross-head motion with an accuracy of 10~$\mu$m. This machine performs both compressive and tensile tests on materials, measuring forces up to 50~kN with an accuracy of {0.5~N}. The mini C-arm consists of a wheeled base, a multiply-hinged arm, and the ``C''-shaped x-ray source/detector structure. Once the base is placed near the materials tester, the hinged arm allows the source and detector to be precisely placed and oriented in 3D to properly image the sample. Finally, the rotation stage attaches rigidly to the Instron, aligned to rotate around the compression axis. The entire setup is surrounded on three sides by a 1~mm lead curtain (Infab Corp) to protect the lab from radiation exposure.

	Figure~\ref{fig:setup1}A depicts the integrated system with all three main components in place for testing. As shown in the figure, the C-arm wraps around the frame of our Instron materials tester. The C-arm is adjusted so that the sample fills up to 90\% of the image width and is centered in the detector's field of view. The detector is then carefully oriented vertically before each experiment, ensuring that the detector plane is perpendicular to the rotation plane. The C-arm is stabilized against motion and sag by securing it to the test bench with aluminum t-slot framing and c-clamps. Between the source and detector, the sample sits fixed to our custom rotation stage.	In this configuration, we can compress a material to measure its stress response, then pause the test, rotate the sample in-place through 360~degrees while acquiring x-ray projections, and finally continue compressing the sample.

	Because the materials tester and x-ray source/detector can be obtained commercially, the key to integration lies in constructing an appropriate rotation stage. The rotation stage must satisfy several requirements for both x-ray tomography \emph{and} material compression. First, the stage must be able to rigidly constrain a sample - up to 150~mm tall and 50~mm in diameter - as it rotates smoothly during imaging. Second, the stage must not obstruct the sample at any point of the rotation. Third, the rotation plane must remain perpendicular to the detector. Fourth, the stage should mate with the material tester and maintain the precise alignment of traditional attachments. Fifth, the stage should sustain up to 10~kN forces achieved when testing strong granular materials. Finally, the stage should not alter or interfere with the measured stress response of the granular materials.

	The stage that we designed to satisfy these requirements is shown and annotated in Fig.~\ref{fig:setup1}B. To satisfy mechanical alignment and loading requirements, all parts were machined to a tolerance of 50~$\mu$m. Moving interfaces were designed with low-friction material pairs, while rigid connections were designed from robust materials (aluminum, steel, acrylic). To allow the sample to move during compression but not during rotation, the sample is attached to two horizontal acrylic plates joined by 3 vertical alignment posts. When compressed, low-friction bushings allow the top plate to slide smoothly along the posts without affecting the measurement. When rotating, the top plate is rigidly fixed to the posts using shaft collars to prevent motion from upward restoring forces. Because the alignment posts will partially obscure the sample when imaging, they are placed 100~mm off-center, so that when rotated, two posts will never align in front of the 50mm-diameter sample. To further minimize reconstruction artifacts from the posts, they are constructed out of carbon fiber rods, which have a low x-ray attenuation cross-section, and are strong enough to hold a compressed sample in place during rotation. Finally, the entire stage is fixed to an aluminum rotation shaft that smoothly spins in brass bushings within the gear bracket. When compressed, the rotation shaft transmits loads to the base of the material tester through a tightened shaft collar resting on the gear bracket, in order to not affect the material measurement.

	\begin{figure}[b]
	\centering
	\includegraphics[width=.9\narrowf]{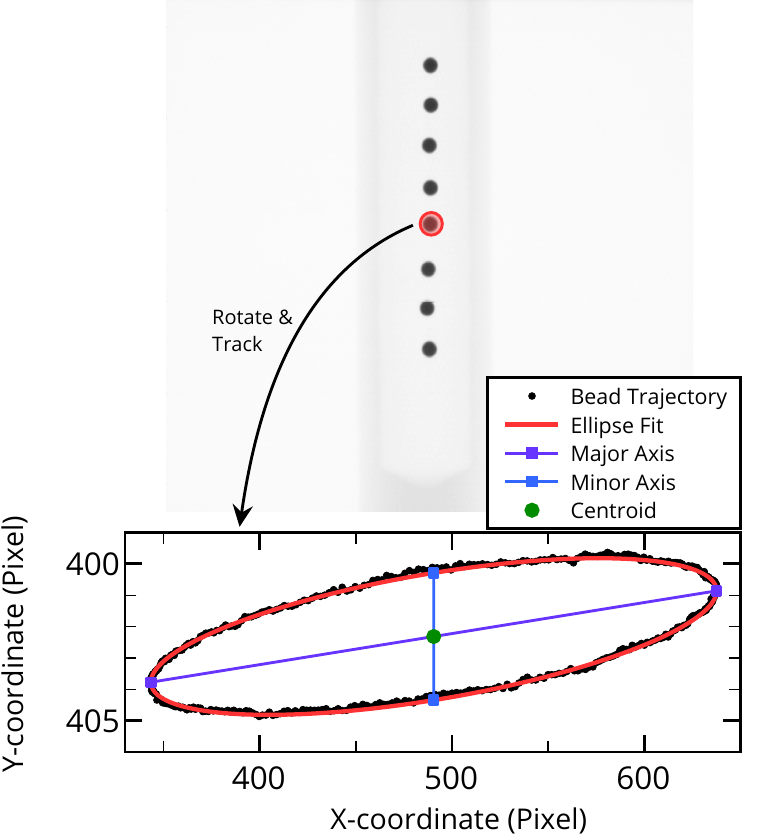}	
	\caption{\label{fig:calibration} Geometric calibration process. We collect projections of 8 steel markers as they are rotated around the rotation stage. A typical projection image is shown in (A). The path that each bead follows can be fitted to an ellipse, as shown in (B) for a single bead. With the fit information for all 8 beads, the system's geometric properties can be uniquely determined.}
	\end{figure}

	The rotation stage is driven by a 200-step microstepping motor (motor part no. 85BYGH450C-03, controller no. CW860-C; Circuit Specialists Inc.), geared down 1:4 to rotate the sample through 800 evenly-spaced projections per 360 degrees. While the stepper motor rotates through consistent steps, we use a rotary encoder (EM1-2-2500-I; US Digital) to guarantee global alignment of the sample between different imaging sequences. This way, we do not have to implement computationally intensive 3D volume registration to directly compare different reconstructed volumes.

	At each of the 800 rotation steps, images are exposed using the C-arm's 80~kVp/100~$\mu$A beryllium x-ray source (50~$\mu$m focal spot diameter). Note that due to software limitations, the source voltage and current are coupled in our Orthoscan system so that increasing the source voltage also increases the source current. The images are acquired with a 15~cm $\times$ 12~cm digital detector, at a resolution of $1024\times968$~pixels$^2$. In order to mitigate beam-hardening artifacts due to polychromatic x-rays, we fix a 1~mm copper plate in front of the source. With a filtered source, experiments reveal that an exposure time of 2~s at a beam energy of 75~kVp produces the best images.

	Both sample rotation and image acquisition are managed using LabView on a dedicated control computer. In order for the control computer to receive images from the C-arm after acquisition, a DICOM server on the control computer must listen for data. Because of its simplicity, we installed the server provided with the DICOM Toolkit.\cite{DCMTK}

	\subsection{Calibration}
	\begin{figure*}[t]
		\centering
		\includegraphics[width=.78\widef]{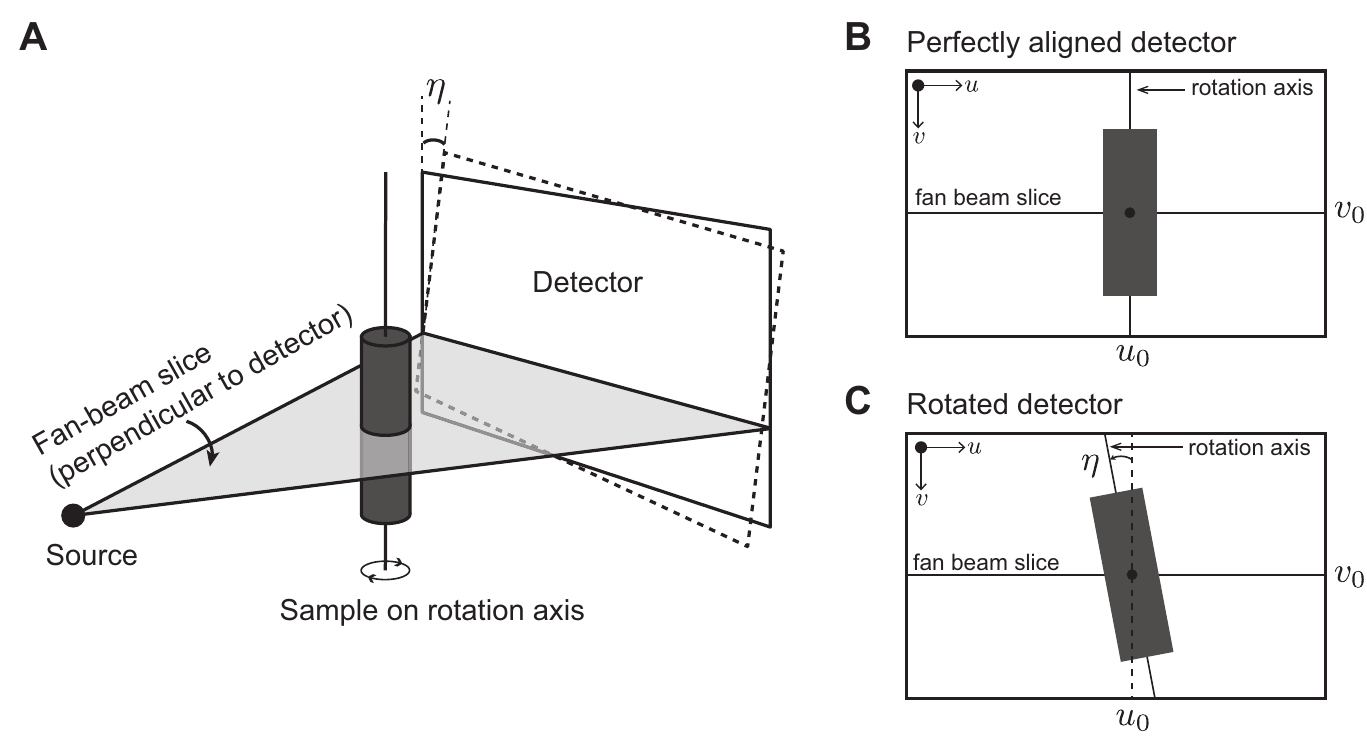}
		\caption{\label{fig:calparams} (A) - Projection geometry for our tomography setup. The fan-beam slice projects as a horizontal line onto the detector. (B) - If the detector is perfectly aligned, then $\eta=0$, the fan-beam slice projects onto the detector at row $v_0$, and the rotation axis projects as a vertical line at column $u_0$. (C) - If the detector is rotated by an angle $\eta>0$, then the rotation axis projects onto the detector at angle $\eta$. In this case, $u_0$ is determined by the $u$-coordinate where the rotation axis intersects the horizontal fan-beam slice at $v_0$.}
	\end{figure*}

	As with all x-ray tube sources, our x-ray source emits a conical beam that projects inhomogeneously onto the detector. Because of this beam geometry, as well as systematic inhomogeneities in the detector, both the system geometry and the image intensity must be calibrated before a sample can be reconstructed. 
	\subsubsection*{Beam Geometry Calibration}
	In order to reconstruct a 3D volume from a set of 2D conebeam projections, certain geometric parameters of the setup need to be measured for each imaging sequence. Before each experiment, we calculate the imaging geometry using the calibration method described in \citeauthor{Yang2006} \cite{Yang2006} This method relies on a calibration phantom, which in our case consists of an acrylic cylinder, with eight 2~mm steel beads spaced evenly along the upper 2/3 of the cylinder. The calibration phantom rigidly attaches  to the rotation stage in the same location as the sample, and is imaged at 400 angles over 360\deg. A projection image of the calibration phantom is shown in Fig.~\ref{fig:calibration}A. After image preprocessing, the 400 calibration images are analyzed to determine the bead centers, yielding a trajectory for each bead. These trajectories form ellipses, whose specific fit parameters provide us with the necessary geometric information \cite{Yang2006}. Fig.~\ref{fig:calibration}B shows a sample bead trajectory and the associated ellipse fit.

	Using this calibration method, we determine five key geometric parameters for our system: the source-detector distance ($R_{SD}$), the source-sample distance ($R_{SI}$), the projection of the rotation axis on the detector($u_0$), the projection of the horizontal fan-beam slice on the detector ($v_0$), and the in-plane detector tilt ($\eta$). Fig.~\ref{fig:calparams} illustrates how the system geometry relates to the parameters $\eta$, $u_0$ and $v_0$. These parameters describe the relationship between the detector, source and the sample, which are together used to define the backprojection geometry as required by the our reconstruction algorithm. For more details, see Section \ref{subsec:recon} as well as \citeauthor{Yang2006}\cite{Yang2006}.

	Typical values of the calibration parameters are shown in Table \ref{tab:calparams}. The parameters will generally vary across experiments if the C-arm is taken down and set up again.

	\begin{table}[h]
		\centering
	  	\renewcommand{\arraystretch}{1.5}
	  	\renewcommand{\tabcolsep}{0.2cm} 
		\begin{tabular}{r|l}
		\hline
		Cal. Param. & Typical Value \\
		\hline
			$R_{SD}$ & 4000~pixels \\
			$R_{SI}$ & 2000-2300~pixels \\
			$u_0$ 	 & col 490 \\
			$v_0$ 	 & row 455 \\
			$\eta$ 	 & 0.010~rad \\
		\hline
		\end{tabular}
		\caption{Typical calibration parameters. $R_{SI}$ varies because we sometimes adjust the source-sample distance to increase resolution at the expense of field of view (see section \ref{sec:resolution} for dependence of resolution on $R_{SI}$).}
		\label{tab:calparams}
	\end{table}

	\subsubsection*{Detector Calibration}

	\begin{figure*}[t]
	\centering
	\includegraphics[width=.98\widef]{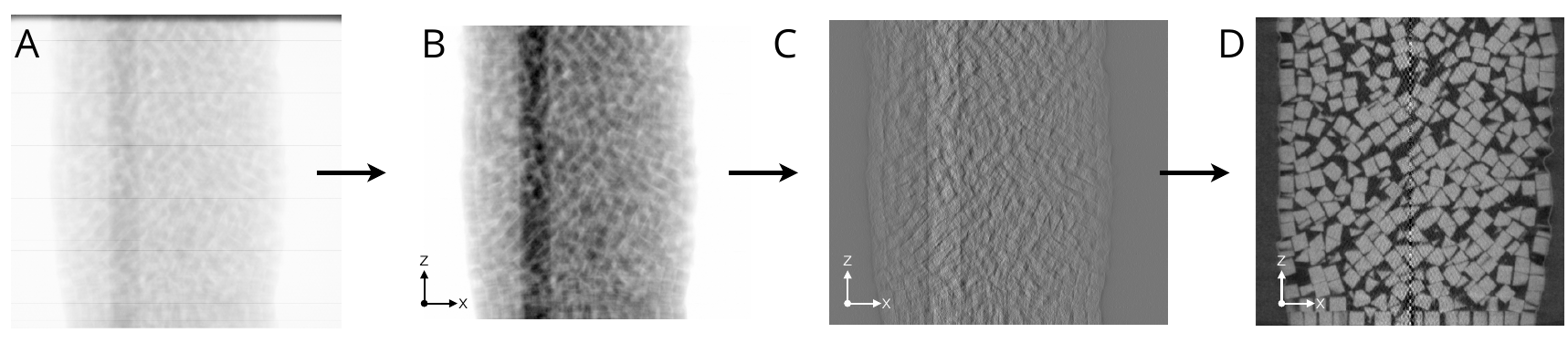}
	\caption{\label{fig:process} To reconstruct a collection of 2D projections into a 3D volume, we process the acquired data through 4 steps. (A) - raw projection images are acquired by the C-arm. (B) - projections are cropped, cleaned, and normalized. (C) - the natural logarithm of the normalized images is compiled into a sinogram ($x$-$z$ slice shown), which is then filtered with a ramp kernel, amplifying higher frequency components within each image. (D) - the filtered sinogram is backprojected into a 3D volume of linear attenuation coefficients using the FDK algorithm\cite{FDKOrig}.}
	\end{figure*}

	To calibrate our reconstruction software for spatial variations in the incident x-ray intensity and detector response, we capture a ``whitefield'' image, or unobstructed x-ray exposure, with which to normalize the projection data. Since this image will be used for normalization, any random noise will typically be amplified and will cause reconstruction artifacts. To minimize random noise, we acquire 50 white-field images, each with the same exposure time and source energy as when imaging the sample. These 50 independent images are then averaged together to create a master white-field image, which is ultimately used to normalize the projection data.

	Additionally, when using a digital detector, electronic noise can produce systematic drift in the detector readings. To minimize this drift, the C-arm is regularly calibrated for ``dark-current correction'' using a built-in calibration tool. Aside from the dark-current correction, no other C-arm calibration is performed with built-in tools; all other image calibration occurs during image preprocessing as described below in section \ref{subsec:preprocess}.

	\subsection{Experimental Procedures}

	In our experiments, we screw the granular packings into the rotation stage, lower the stage's top-plate onto the sample, and screw a threaded rod into the top of the sample through a hole in the top plate. This threaded rod is then tightly gripped by the material tester's wedge-grip accessory so that the sample is rigidly connected to the material tester's load cell and cross-head. During compression, the cross-head lowers quasi-statically, at a rate of 8~mm/min. At pre-determined compression levels, the cross-head pauses so that the sample can be imaged. 

	Before imaging, the sample is fixed at its current degree of compression by tightening shaft collars onto the carbon fiber alignment rods. After the sample is fixed, the gripper is disconnected from the threaded rod attached to the sample, so that the rotation stage can rotate freely. The sample is then imaged at 0.47\deg~increments, for a total of 800 images per 360\deg. While cone beam reconstruction can be accomplished by imaging over a smaller angular range, we observed fewer reconstruction artifacts when reconstructing with data acquired over 360\deg. 

	At each angle, the x-ray source and detector expose the sample for 2~seconds and then wait for 1.5~seconds while the sample is rotated. The 1.5~second break is necessary to ensure that the C-arm has time to save the previous image and process the next activation signal. The imaging process is computer-controlled through LabView, and signals are sequentially sent to the motor controller and C-arm using analog signals from a National Instruments DAQ. In order to trigger the C-arm from the computer, we pass a square pulse through an optocoupler that closes the external triggering circuit on the C-arm. Software limitations on the C-arm require that each set of 800 images is broken up into two runs of 400 images each. As a result, our system requires brief operator interaction halfway through the imaging process.

	Once imaging at one compression level is complete, we reattach the material tester to the sample and resume compressing the sample.	After an experiment, we retrieve the image data from the C-arm using a built-in DICOM export feature. The internal C-arm export software contains a bug that occasionally distorts exported images. This distortion consistently takes the form of a vertical displacement of the image and black striping near the bottom. Leveraging the consistency of the bug, the receiving server checks for corrupted DICOM files and rejects them, alerting the operator to manually re-export the malformed images.
	\section{3D Reconstruction}
	\subsection{Image Preprocessing}\label{subsec:preprocess}
	Once properly exported, output images require preprocessing to correct for two disruptive image artifacts. The first artifact, dead detector pixels, introduces black lines and points into the acquired images. We identify dead pixels using a local threshold, and interpolate their values using an inpainting algorithm \cite{Criminisi:Inpainting}. The specific inpainting algorithm we use is available in the OpenCV computer vision library \cite{opencv_library}.

	The second artifact, non-uniform illumination, is caused separately by both the x-ray source and detector. First, inhomogeneities in the x-ray source produce spatial variations in the x-ray intensity at different points on the detector. Second, the detector housing contributes a feint hexagonal grid to the images. In order to remove both of these variations from the raw images, we normalize projections of the sample by a white-field image, which is taken under the same conditions as the sample, but without any obstructions in the field of view. Note - such a normalization step naturally arises when reconstructing tomographic data \cite{KakSlaney}.

	\begin{figure*}[!t]
	\centering
	\includegraphics[width=.9\widef]{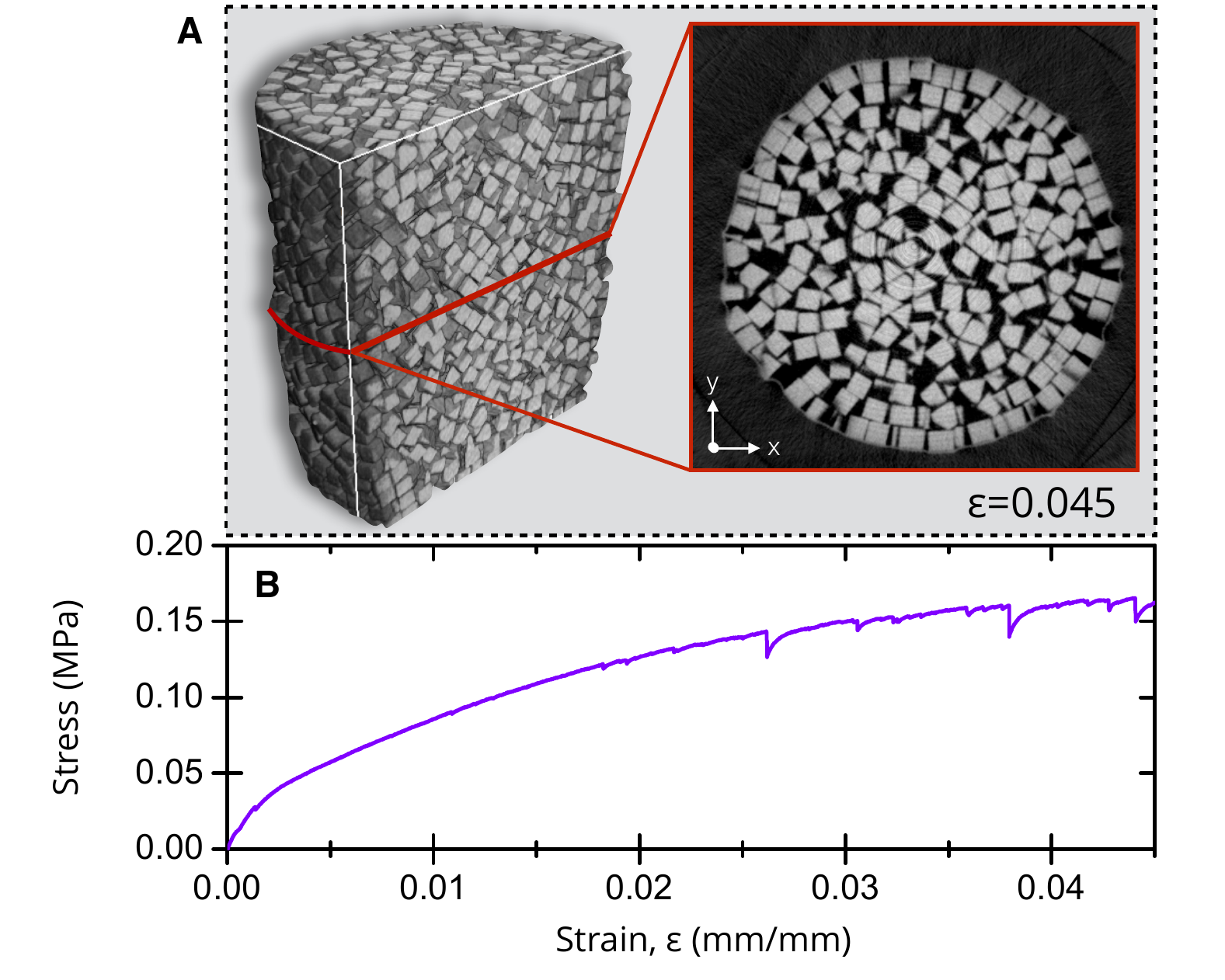}
	\caption{\label{fig:recon} (A) 3D rendering and 2D cross section of reconstruction data from a granular packing of cubes after 4.5\% compression. The corresponding stress-strain curve for that experimental run is shown in (B).}
	\end{figure*}

	\vspace{-1em}
	\subsection{Tomographic Reconstruction}\label{subsec:recon}
	The purpose of the reconstruction process is to back out the 3D map of linear attenuation coefficients within the sample. Our reconstruction software is a custom implementation of the Feldkamp-Davis-Kress (FDK) cone-beam reconstruction algorithm \cite{FDKOrig}. The software takes a preprocessed image stack and the calibration information, and reconstructs a 3D volume of attenuation coefficients. Our code is designed specially to run in parallel on a CUDA-enabled GPU. We used a GeForce GTX 550 Ti graphics card (192 CUDA Cores) on a custom computer with 16~GB RAM, a 3.3GHz Intel i5 processor, running Ubuntu Linux. With this computing configuration, we are able to process and reconstruct full volumes in under 1~hour.

	All of our software for preprocessing and reconstruction is implemented in Python and C. In order to efficiently handle the images and volume data, we use many open-source Python libraries designed for scientific computing \cite{SciPyNumPy,matplotlib,mahotas,opencv_library}.

	The reconstruction algorithm was implemented following Kak and Slaney, Ch. 3\cite{KakSlaney}, without modification. To reconstruct the map of attenuation coefficients within the sample, the preprocessed images are assembled into a stack of projection image known as a sinogram. In the process of assembly, the natural logarithm of the data is taken, so that the sinogram represents $\ln(I/I_0)$ where $I_0$ is the whitefield intensity. The sinogram is then filtered with a ramp kernel \cite{KakSlaney} and backprojected into a 3D volume using the FDK reconstruction algorithm \cite{FDKOrig}. Fig.~\ref{fig:process} demonstrates the process that takes a series of projection images into a final reconstruction. As shown in the figure, the carbon-fiber posts appear in the processed images. At first glance, these posts may seem to threaten the quality of the reconstructed volumes because the posts partially obstruct the sample. However, as shown in Fig.~\ref{fig:process}D, the visible effect that the posts have on the reconstruction image is to introduce black striping outside of the packing. This striping arises because the posts briefly exit the field of view while the packing is rotated. As a result, the posts introduce a low-spatial-frequency artifact into the reconstruction, which corresponds to the missing information at the post position when the posts exit the detector's field of view. Since the posts rotate outside of the horizontal field-of-view, this low-frequency artifact only reconstructs to the corners of the image where the distance to the image center exceeds the horizontal image size. In addition to this low-frequency artifact, the posts also introduce some mid and high-frequency noise into the reconstruction. These effects are incorporated into our empirical noise estimates discussed below.

	We imaged a packing of 2~mm plastic cubes after triaxial compression, and present the reconstructed packing along with the stress response in Fig.~\ref{fig:recon}. Fig.~\ref{fig:recon}A depicts a rendering of the reconstructed volume, partially cut to reveal the internal structure. The second image shows an $x-y$ slice of the reconstructed packing. Fig.~\ref{fig:recon}B presents the compressive response of the cube packing, measured before imaging.

	\begin{figure}[!t]
	\centering
	\includegraphics[width=\narrowf]{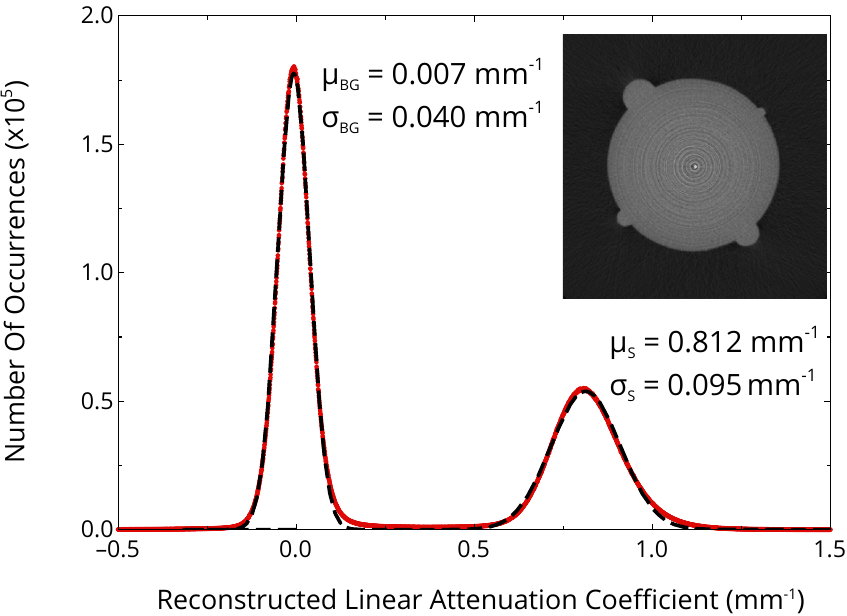}
	\caption{\label{fig:grayhist} Histogram of the reconstructed voxel values in a reconstruction of a 3D-printed phantom. The two peaks reflect the density of air and the plastic respectively, which are both corrupted by noise. By approximating the noise as Gaussian, we fit the histogram to extract the peak centers and widths in order to calculate the image quality in our reconstructions. The red dots indicate the calculated histogram values, while the dashed line represents a two-peak Gaussian fit to the data. The inset image depicts an $x-y$ slice of the reconstruction.}
	\end{figure}

	\subsection{Reconstruction Quality}

	Our reconstruction quality can be assessed in three parts: noise, {spatial resolution and artifacts}.

	\subsubsection*{Noise}

	The signal quality of our reconstructions can be quantified by the signal-to-noise ratio (SNR), which captures our ability to distinguish a reconstructed sample from background noise. To measure SNR, we reconstruct a homogeneous 3D-printed phantom, which is shown as an inset to Fig.~\ref{fig:grayhist}. The figure shows the histogram of reconstructed gray values in the 3D volume, demonstrating two clear peaks near 0 and 0.81 (mm$^{-1}$). The first peak corresponds to the air (background) while the second peak corresponds to the 3D-printed phantom. To quantify the image characteristics, we model the plastic phantom and the air as well-defined by a single grayscale value corrupted by Gaussian noise, and fit to find the mean and standard deviation of each peak. While the fit reveals that neither peak is perfectly Gaussian, it suffices to estimate signal quality.

	We calculate a common differential SNR, comparing the contrast between the 3D-printed phantom and the air to the average width of the noise distribution within the two media\cite{PrinceLinks}: $$ \text{SNR} = \sqrt{2}\frac{\mu_S-\mu_{BG}}{\sqrt{\sigma_S^2 + \sigma_{BG}^2}} = 11.0.$$

	This SNR indicates that we have fairly high contrast allowing us to statistically distinguish the 3D-printed material from the air despite the moderate noise. A smart processing algorithm could be implemented to properly segment the sample from the background, accounting for the measured noise distribution.

	Note that some negative values appear in our reconstructed images. Negative values can arise in filtered backprojection reconstructions because the filtration kernel introduces negative values into the filtered projections. For perfectly acquired data, these negatives cancel out during the backprojection process, but in the presence of noise and undersampling, some negative values can survive in the final reconstructed image, especially in low attenuation regions (see Kak and Slaney, Ch. 5)\cite{KakSlaney}. Indeed, our reconstructions only obtain negative values in background regions, as demonstrated in plots \ref{fig:resolution}B and \ref{fig:artifacts}B.

	\subsubsection*{Spatial Resolution}\label{sec:resolution}

	We reconstruct volumes of 400$\times$400$\times$500 voxels, with a voxel edge length of 150~$\mu$m. For our needs, these reconstruction parameters sufficiently balanced resolution and computation time. However, our imaging hardware is theoretically limited only by the effective focal spot and detector pixel sizes, as scaled to the rotation axis. In this section, by assuming a 2D fan-beam geometry,  we propagate these two limiting factors through the reconstruction process in order to estimate the 3D resolution of our imaging system.

	\begin{figure*}[!t]
	\centering
	\includegraphics[width=.8\widef]{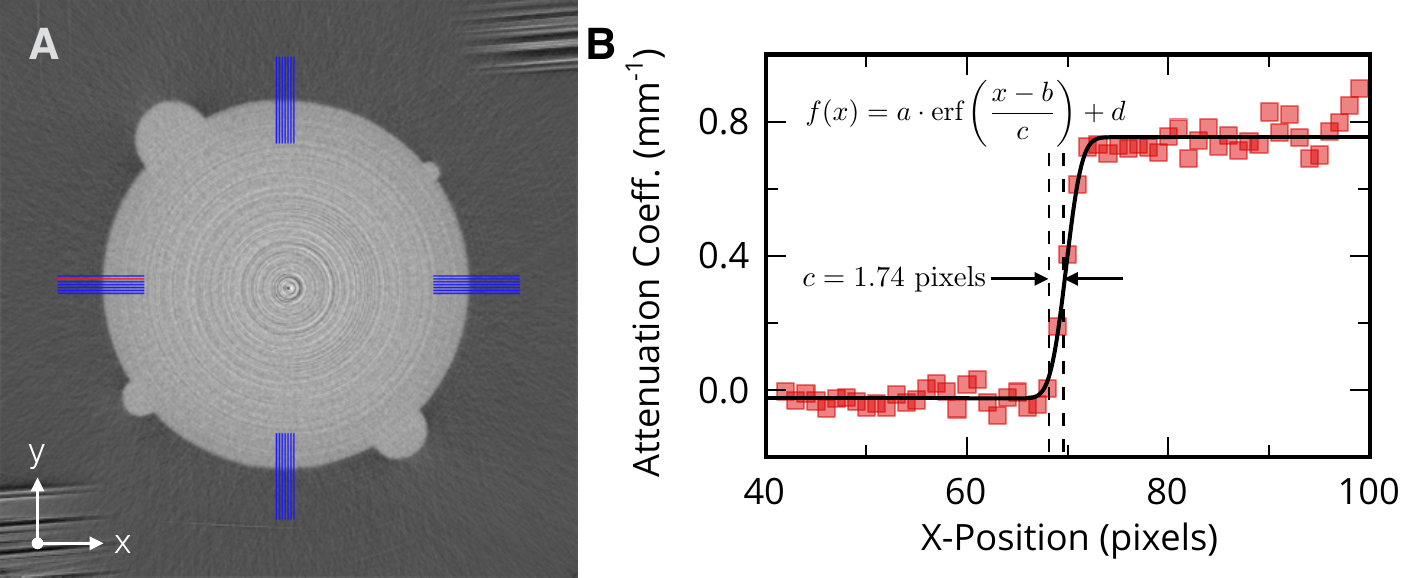}
	\caption{\label{fig:resolution} To measure our system's spatial resolution, we reconstruct the phantom shown in (A), and extract 28 edge profiles from different points around the perimeter (highlighted in blue and red). Fitting each profile to an error function, we can extract the width parameter, $c$, and relate this number to the resolution of our system. In (B), the red profile from (A) is plotted. Red squares represent the measured intensity values along the profile, and the black line denotes the error function fit.}	
	\end{figure*}

	Based on the reconstruction algorithm and beam geometry, the final reconstructed image resolution will depend on the projection resolution \emph{at the sample}. This means that the finite detector resolution (pixel spacing) must be geometrically projected to the sample center to characterize the resolution of the system. Additionally, the resolution can be affected because the x-ray beam emanates from a finite focal spot. Thus, to estimate the overall resolution in projection images, we combine the resolution restrictions imposed by finite detector pixel spacing and by finite source width. In the case of finite detector pixel spacing, $du=150\mu$m, we assume a point source and can consider the geometric magnification between the source and detector. If the source-detector spacing is $R_{SD}$ and the source-sample distance is $R_{SI}$, then the spatial resolution at the sample center is demagnified to $du' = \frac{R_{SI}}{R_{SD}} du.$ To account for finite source width, $ds=50\mu$m, we assume infinitesimal detector resolution and can similarly demagnify the source resolution onto the sample plane, $ds'=\frac{R_{SD}-R_{SI}}{R_{SD}} ds.$ Combining these two contributions, we estimate that the spatial resolution of our projections at the sample is $$dx = \sqrt{du'^2 + ds'^2} \approx 80\mu\text{m}.$$

	\noindent The projected spatial resolution, $dx$, can then be propagated through the reconstruction process to estimate the maximum spatial resolution in the reconstructed volumes. At the very least, the projected resolution must be combined in quadrature with the reconstruction voxel size. For $150\mu$m voxels, this yields a best case reconstruction resolution of $175\mu$m.
 
	In order to quantify the resolution of our hardware and reconstruction algorithm directly, we image a resolution phantom shown in Fig.~\ref{fig:resolution}A. The phantom is 3D-printed with a resolution of 30~$\mu$m, so that surface roughness is not detectable with our imaging setup. To measure the spatial resolution of our imaging system, we first reconstruct the phantom into a volume as described above. We then measure the edge profile at 28 locations around the sample (highlighted in Fig.~\ref{fig:resolution}A). A sample profile along the left edge of the phantom is shown in Fig.~\ref{fig:resolution}B. 

	In an ideal system, we would expect the reconstructed radial profile to reflect a clean step function at the edge of the phantom. However, since our imaging system has a finite impulse response, this discontinuous edge will be blurred by an approximately Gaussian point spread function (PSF), producing an error function edge profile. Mathematically, then, we expect the edge profile in the reconstruction to be of the form $$ f(x) = a\cdot\text{erf}\left(\frac{x-b}{c}\right) + d,$$ which is described by four free parameters. Of these parameters, the width, $c$, also describes the standard deviation of the Gaussian PSF, where {$c =\sqrt 2 \sigma_{gauss}.$} As shown in Fig.~\ref{fig:resolution}B, we fit the profiles to an error function, and used the fitted $c$ to calculate our resolution as $$\delta = (FWHM)_{Gaussian} = c\cdot 2\sqrt{\ln 2}.$$

	\noindent This resolution criterion captures the ability of our imaging system to identify edges in a sample. Generally, a complete understanding of the system resolution can be obtained by calculating the Modulation Transfer Function (MTF) \cite{Hsieh2009}. The MTF is calculated as magnitude of the Fourier Transform of the system's Point Spread Function (PSF), which we assume is Gaussian. For some analysis tasks, a specific point (such as 5\% contrast) on the MTF is employed as the useful resolution criterion \cite{Hsieh2009}. However, this criterion is highly application-specific and can suffer drawbacks when used in other applications \cite{Hsieh2009}. Instead, since our ultimate task is to distinguish closely-packed particles by their high-contrast edges, the blurred edge width is a more appropriate resolution criterion to employ.

	Averaging across 28 independent edge profiles to minimize the effects of reconstruction noise, we calculate our imaging resolution to be ${\bar\delta=3.4\pm0.5}$~pixels. For our reconstructed voxel edge length of 150~$\mu$m, this translates into a spatial resolution of $\bar\delta=520\mu$m. Since $\bar\delta$ is higher than our estimated hardware limit, a more complex PSF is clearly arising from the reconstruction process.

	When considered as the width of the PSF, $\bar\delta$ sheds light on the origin of the wide signal peak in Fig.~\ref{fig:grayhist} - not only does noise contribute to this peak, but a wider PSF significantly widens the peak as well! Stemming from this observation, if this reconstruction PSF can be accurately measured, then the resolution \emph{and} SNR can be improved by using advanced deconvolution techniques.

	\subsubsection*{Artifacts}

\begin{figure}[!t]
	\centering
	\includegraphics[width=\narrowf]{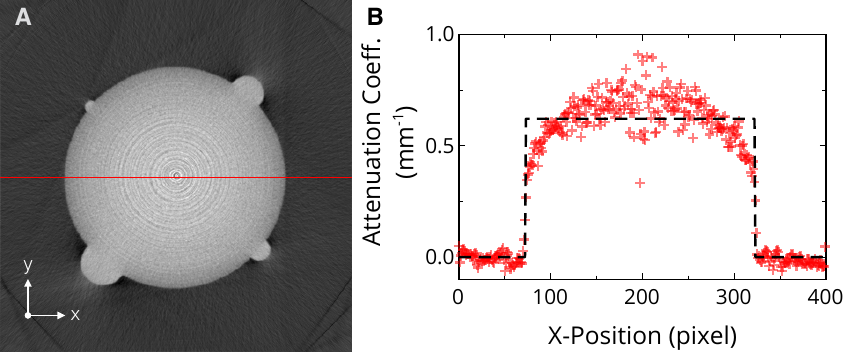}
	\caption{\label{fig:artifacts} Three common artifacts that we encounter are rings, capping and beam-hardening. In (A), the reconstructed slice clearly shows the concentric ringing artifact. Although the ringing artifact does not visually obstruct the homogeneous phantom, the rings disrupt image analysis of random packings. Additionally, the dark shadows and bright fans near the edge of the phantom arise from beam hardening during acquisition. In (B), a cross section of the image in (A) demonstrates the capping artifact clearly: while we expect the cylindrical phantom to appear as a clean step in the cross-section plot (black dotted line), the reconstructed profile (red +) rises nonlinearly, achieving maximum brightness at the center of the image.}
	\end{figure}

	Despite our high-quality reconstructions, we encounter three common artifacts in our imaging data: rings, capping artifacts, and beam-hardening artifacts. Figure~\ref{fig:artifacts} demonstrates these three effects clearly. 

	Ring artifacts manifest themselves in every $z$-slice of the reconstruction, with some slices demonstrating the artifact worse than others. These rings can appear for different reasons, but are most commonly caused in our system by improperly normalized bad detector pixels. Along with degrading the SNR of the system, they introduce hard edges into the reconstruction that obstruct certain image analysis techniques based on gradient detection. Fortunately, ring artifacts can be reduced in many ways including, better x-ray detectors, better bad-pixel inpainting algorithms, or iterative reconstruction techniques that can ignore bad detector pixels in the reconstruction process. During image analysis, the rings contribute to high-frequency noise and can often be reduced using a local filter such as a Gaussian or median filter.

	In some cases, our reconstructions also demonstrate capping artifacts. Such artifacts cause a homogeneous material to appear brighter near the image center and dimmer near the edges. This artifact widens the image histogram, reduces the SNR, and complicates analysis of the packing. Typically, capping artifacts are caused by overexposed (and saturated) projection images \cite{Reiser2009}. In our system, we observed that capping artifacts were reduced by reducing the beam energy, consistent with reports that this artifact arises from overexposure. Possible fixes for this artifact could include exposure optimization before projection images are acquired, or digital overexposure correction before or during reconstruction.

	Additionally, some of our reconstructions contain mild beam-hardening artifacts. Such artifacts arise when low-energy x-rays from a polychromatic source are attenuated by the sample more than the higher-energy x-rays. Beam-hardening is visible in Fig.~\ref{fig:artifacts}A at the edges of the phantom as well as in Fig.~\ref{fig:recon} at the edges of the packing. This artifact can be substantially reduced by placing a copper screening plate in front of the x-ray source as well as appropriately adjusting the exposure settings for projection images.

	Very rarely, we have encountered other, more complex artifacts in the reconstruction process. Most of the time, these artifacts arise from detector motion during the experiment. If the C-arm is not properly secured, it can sag under gravity and can also shift if vibrated or bumped, changing the system geometry from the initial calibration state. Since the reconstruction process relies heavily on the geometric calibration, even small changes to the source/detector geometry can be enough to introduce significant artifacts into the reconstruction.

	\section{Conclusion}

	In this paper, we demonstrate how a medical mini C-arm can be integrated with a traditional materials testing apparatus to perform x-ray tomography of a material sample while it is being mechanically stressed. Our system provides a number of advantages over existing non-invasive measurement techniques for materials testing. Our system:
	\begin{itemize}
		\item is low-cost compared to existing medical and synchrotron imaging solutions
		\item can be integrated with existing materials testing experiments
		\item provides a large field of view compatible with many typical soft matter experiments
	\end{itemize}

	Because it is built around x-ray tomography, this system can be applied much more broadly than to the granular packings presented here. Many typical polymer-based rubbers and plastics used in current soft-matter experiments have ideal x-ray cross sections for reconstruction with our system. Additionally, with the recent explosion of 3D-printing, more complex multi-material systems can be fabricated easily for testing in a system such as ours.

	By expanding to different material classes, such a system, paired with innovative 3D image analysis, can help to quantify the relationship between the global and local responses in many different microstructural materials. Many phenomena naturally arise as candidates for this kind of testing: buckling modes of 3D foams and networks, and failure identification in 3D materials represent a small subset of the materials problems that would benefit from integrated measurements such as ours.	Ultimately, integrated stress and microstructure measurements enable quantitative analysis of the relationship between local material behavior and global response.

	\subsection*{Acknowledgements}
	The authors would like to thank Peter Eng for constructive feedback on the initial design, Helmut Krebs for his guidance building the rotation stage, M. Schr\"oter for suggesting we use a copper plate in front of the source, K. Murphy and M. Miskin for valuable discussions, and the UChicago Office of Radiation Safety for their assistance while setting up this experiment. This work was supported by the US Army Research Office through grant W911NF-12-1-0182.

	\subsection*{References}
	\bibliography{granx}

\begin{thebibliography}{48}%
\makeatletter
\providecommand \@ifxundefined [1]{%
 \@ifx{#1\undefined}
}%
\providecommand \@ifnum [1]{%
 \ifnum #1\expandafter \@firstoftwo
 \else \expandafter \@secondoftwo
 \fi
}%
\providecommand \@ifx [1]{%
 \ifx #1\expandafter \@firstoftwo
 \else \expandafter \@secondoftwo
 \fi
}%
\providecommand \natexlab [1]{#1}%
\providecommand \enquote  [1]{``#1''}%
\providecommand \bibnamefont  [1]{#1}%
\providecommand \bibfnamefont [1]{#1}%
\providecommand \citenamefont [1]{#1}%
\providecommand \href@noop [0]{\@secondoftwo}%
\providecommand \href [0]{\begingroup \@sanitize@url \@href}%
\providecommand \@href[1]{\@@startlink{#1}\@@href}%
\providecommand \@@href[1]{\endgroup#1\@@endlink}%
\providecommand \@sanitize@url [0]{\catcode `\\12\catcode `\$12\catcode
  `\&12\catcode `\#12\catcode `\^12\catcode `\_12\catcode `\%12\relax}%
\providecommand \@@startlink[1]{}%
\providecommand \@@endlink[0]{}%
\providecommand \url  [0]{\begingroup\@sanitize@url \@url }%
\providecommand \@url [1]{\endgroup\@href {#1}{\urlprefix }}%
\providecommand \urlprefix  [0]{URL }%
\providecommand \Eprint [0]{\href }%
\providecommand \doibase [0]{http://dx.doi.org/}%
\providecommand \selectlanguage [0]{\@gobble}%
\providecommand \bibinfo  [0]{\@secondoftwo}%
\providecommand \bibfield  [0]{\@secondoftwo}%
\providecommand \translation [1]{[#1]}%
\providecommand \BibitemOpen [0]{}%
\providecommand \bibitemStop [0]{}%
\providecommand \bibitemNoStop [0]{.\EOS\space}%
\providecommand \EOS [0]{\spacefactor3000\relax}%
\providecommand \BibitemShut  [1]{\csname bibitem#1\endcsname}%
\let\auto@bib@innerbib\@empty
\bibitem [{\citenamefont {Duran}(1999)}]{Duran1999}%
  \BibitemOpen
  \bibfield  {author} {\bibinfo {author} {\bibfnamefont {J.}~\bibnamefont
  {Duran}},\ }\href@noop {} {\emph {\bibinfo {title} {Sands, powders, and
  grains : an introduction to the physics of granular materials}}}\ (\bibinfo
  {publisher} {Springer},\ \bibinfo {address} {Berlin},\ \bibinfo {year}
  {1999})\BibitemShut {NoStop}%
\bibitem [{\citenamefont {Andreotti}, \citenamefont {Forterre},\ and\
  \citenamefont {Pouliquen}(2013)}]{Andreotti2013}%
  \BibitemOpen
  \bibfield  {author} {\bibinfo {author} {\bibfnamefont {B.}~\bibnamefont
  {Andreotti}}, \bibinfo {author} {\bibfnamefont {Y.}~\bibnamefont {Forterre}},
  \ and\ \bibinfo {author} {\bibfnamefont {O.}~\bibnamefont {Pouliquen}},\
  }\href@noop {} {\emph {\bibinfo {title} {Granular media between fluid and
  solid}}}\ (\bibinfo  {publisher} {Cambridge University Press},\ \bibinfo
  {address} {Cambridge},\ \bibinfo {year} {2013})\BibitemShut {NoStop}%
\bibitem [{\citenamefont {Jaeger}, \citenamefont {Nagel},\ and\ \citenamefont
  {Behringer}(1996)}]{Jaeger1996}%
  \BibitemOpen
  \bibfield  {author} {\bibinfo {author} {\bibfnamefont {H.~M.}\ \bibnamefont
  {Jaeger}}, \bibinfo {author} {\bibfnamefont {S.~R.}\ \bibnamefont {Nagel}}, \
  and\ \bibinfo {author} {\bibfnamefont {R.~P.}\ \bibnamefont {Behringer}},\
  }\href@noop {} {\bibfield  {journal} {\bibinfo  {journal} {Reviews of Modern
  Physics}\ }\textbf {\bibinfo {volume} {68}},\ \bibinfo {pages} {1259}
  (\bibinfo {year} {1996})}\BibitemShut {NoStop}%
\bibitem [{\citenamefont {Herrmann}, \citenamefont {Hovi},\ and\ \citenamefont
  {Luding}(1998)}]{Herrmann1998}%
  \BibitemOpen
  \bibfield  {author} {\bibinfo {author} {\bibfnamefont {H.}~\bibnamefont
  {Herrmann}}, \bibinfo {author} {\bibfnamefont {J.}~\bibnamefont {Hovi}}, \
  and\ \bibinfo {author} {\bibfnamefont {S.}~\bibnamefont {Luding}},\ }\href
  {http://books.google.com/books?id=TZyOKD8bJRgC} {\emph {\bibinfo {title}
  {Physics of Dry Granular Media}}},\ NATO Advanced Science Institutes Series.
  Series E, Applied Sciences\ (\bibinfo  {publisher} {Springer},\ \bibinfo
  {year} {1998})\BibitemShut {NoStop}%
\bibitem [{\citenamefont {Liu}\ \emph {et~al.}(1995)\citenamefont {Liu},
  \citenamefont {Nagel}, \citenamefont {Schecter}, \citenamefont {Coppersmith},
  \citenamefont {Majumdar}, \citenamefont {Narayan},\ and\ \citenamefont
  {Witten}}]{Liu1995}%
  \BibitemOpen
  \bibfield  {author} {\bibinfo {author} {\bibfnamefont {C.~H.}\ \bibnamefont
  {Liu}}, \bibinfo {author} {\bibfnamefont {S.~R.}\ \bibnamefont {Nagel}},
  \bibinfo {author} {\bibfnamefont {D.~A.}\ \bibnamefont {Schecter}}, \bibinfo
  {author} {\bibfnamefont {S.~N.}\ \bibnamefont {Coppersmith}}, \bibinfo
  {author} {\bibfnamefont {S.}~\bibnamefont {Majumdar}}, \bibinfo {author}
  {\bibfnamefont {O.}~\bibnamefont {Narayan}}, \ and\ \bibinfo {author}
  {\bibfnamefont {T.~A.}\ \bibnamefont {Witten}},\ }\href@noop {} {\bibfield
  {journal} {\bibinfo  {journal} {Science}\ }\textbf {\bibinfo {volume}
  {269}},\ \bibinfo {pages} {513} (\bibinfo {year} {1995})}\BibitemShut
  {NoStop}%
\bibitem [{\citenamefont {Brujic}\ \emph {et~al.}(2003)\citenamefont {Brujic},
  \citenamefont {Edwards}, \citenamefont {Grinev}, \citenamefont {Hopkinson},
  \citenamefont {Brujic},\ and\ \citenamefont {Makse}}]{Brujic2003}%
  \BibitemOpen
  \bibfield  {author} {\bibinfo {author} {\bibfnamefont {J.}~\bibnamefont
  {Brujic}}, \bibinfo {author} {\bibfnamefont {S.~F.}\ \bibnamefont {Edwards}},
  \bibinfo {author} {\bibfnamefont {D.~V.}\ \bibnamefont {Grinev}}, \bibinfo
  {author} {\bibfnamefont {I.}~\bibnamefont {Hopkinson}}, \bibinfo {author}
  {\bibfnamefont {D.}~\bibnamefont {Brujic}}, \ and\ \bibinfo {author}
  {\bibfnamefont {H.~A.}\ \bibnamefont {Makse}},\ }\href@noop {} {\bibfield
  {journal} {\bibinfo  {journal} {Faraday Discussions}\ }\textbf {\bibinfo
  {volume} {123}},\ \bibinfo {pages} {207} (\bibinfo {year}
  {2003})}\BibitemShut {NoStop}%
\bibitem [{\citenamefont {Dijksman}\ \emph {et~al.}(2012)\citenamefont
  {Dijksman}, \citenamefont {Rietz}, \citenamefont {L{\H{o}}rincz},
  \citenamefont {van Hecke},\ and\ \citenamefont {Losert}}]{Dijksman2012}%
  \BibitemOpen
  \bibfield  {author} {\bibinfo {author} {\bibfnamefont {J.~A.}\ \bibnamefont
  {Dijksman}}, \bibinfo {author} {\bibfnamefont {F.}~\bibnamefont {Rietz}},
  \bibinfo {author} {\bibfnamefont {K.~A.}\ \bibnamefont {L{\H{o}}rincz}},
  \bibinfo {author} {\bibfnamefont {M.}~\bibnamefont {van Hecke}}, \ and\
  \bibinfo {author} {\bibfnamefont {W.}~\bibnamefont {Losert}},\ }\href@noop {}
  {\bibfield  {journal} {\bibinfo  {journal} {Review of Scientific
  Instruments}\ }\textbf {\bibinfo {volume} {83}},\ \bibinfo {pages} {11301}
  (\bibinfo {year} {2012})}\BibitemShut {NoStop}%
\bibitem [{\citenamefont {Nakagawa}\ \emph {et~al.}(1993)\citenamefont
  {Nakagawa}, \citenamefont {Altobelli}, \citenamefont {Caprihan},
  \citenamefont {Fukushima},\ and\ \citenamefont {Jeong}}]{Nakagawa1993}%
  \BibitemOpen
  \bibfield  {author} {\bibinfo {author} {\bibfnamefont {M.}~\bibnamefont
  {Nakagawa}}, \bibinfo {author} {\bibfnamefont {S.~A.}\ \bibnamefont
  {Altobelli}}, \bibinfo {author} {\bibfnamefont {A.}~\bibnamefont {Caprihan}},
  \bibinfo {author} {\bibfnamefont {E.}~\bibnamefont {Fukushima}}, \ and\
  \bibinfo {author} {\bibfnamefont {E.~K.}\ \bibnamefont {Jeong}},\ }\href@noop
  {} {\bibfield  {journal} {\bibinfo  {journal} {Experiments in Fluids}\
  }\textbf {\bibinfo {volume} {16}},\ \bibinfo {pages} {54} (\bibinfo {year}
  {1993})}\BibitemShut {NoStop}%
\bibitem [{\citenamefont {Kuperman}(1996)}]{Kuperman1996}%
  \BibitemOpen
  \bibfield  {author} {\bibinfo {author} {\bibfnamefont {V.~Y.}\ \bibnamefont
  {Kuperman}},\ }\href@noop {} {\bibfield  {journal} {\bibinfo  {journal}
  {Physical Review Letters}\ }\textbf {\bibinfo {volume} {77}},\ \bibinfo
  {pages} {1178} (\bibinfo {year} {1996})}\BibitemShut {NoStop}%
\bibitem [{\citenamefont {Hill}, \citenamefont {Caprihan},\ and\ \citenamefont
  {Kakalios}(1997)}]{Hill1997}%
  \BibitemOpen
  \bibfield  {author} {\bibinfo {author} {\bibfnamefont {K.~M.}\ \bibnamefont
  {Hill}}, \bibinfo {author} {\bibfnamefont {A.}~\bibnamefont {Caprihan}}, \
  and\ \bibinfo {author} {\bibfnamefont {J.}~\bibnamefont {Kakalios}},\
  }\href@noop {} {\bibfield  {journal} {\bibinfo  {journal} {Physical Review
  Letters}\ }\textbf {\bibinfo {volume} {78}},\ \bibinfo {pages} {50} (\bibinfo
  {year} {1997})}\BibitemShut {NoStop}%
\bibitem [{\citenamefont {Nakagawa}\ \emph {et~al.}(1997)\citenamefont
  {Nakagawa}, \citenamefont {Altobelli}, \citenamefont {Caprihan},\ and\
  \citenamefont {Fukushima}}]{Nakagawa1997}%
  \BibitemOpen
  \bibfield  {author} {\bibinfo {author} {\bibfnamefont {M.}~\bibnamefont
  {Nakagawa}}, \bibinfo {author} {\bibfnamefont {S.~A.}\ \bibnamefont
  {Altobelli}}, \bibinfo {author} {\bibfnamefont {A.}~\bibnamefont {Caprihan}},
  \ and\ \bibinfo {author} {\bibfnamefont {E.}~\bibnamefont {Fukushima}},\
  }\href@noop {} {\bibfield  {journal} {\bibinfo  {journal} {Chemical
  Engineering Science}\ }\textbf {\bibinfo {volume} {52}},\ \bibinfo {pages}
  {4423} (\bibinfo {year} {1997})}\BibitemShut {NoStop}%
\bibitem [{\citenamefont {Oda}\ and\ \citenamefont {Kazama}(1998)}]{Oda1998}%
  \BibitemOpen
  \bibfield  {author} {\bibinfo {author} {\bibfnamefont {M.}~\bibnamefont
  {Oda}}\ and\ \bibinfo {author} {\bibfnamefont {H.}~\bibnamefont {Kazama}},\
  }\href@noop {} {\bibfield  {journal} {\bibinfo  {journal} {Geotechnique}\
  }\textbf {\bibinfo {volume} {48}},\ \bibinfo {pages} {465} (\bibinfo {year}
  {1998})}\BibitemShut {NoStop}%
\bibitem [{\citenamefont {Seymour}\ \emph {et~al.}(2000)\citenamefont
  {Seymour}, \citenamefont {Caprihan}, \citenamefont {Altobelli},\ and\
  \citenamefont {Fukushima}}]{Seymour2000}%
  \BibitemOpen
  \bibfield  {author} {\bibinfo {author} {\bibfnamefont {J.~D.}\ \bibnamefont
  {Seymour}}, \bibinfo {author} {\bibfnamefont {A.}~\bibnamefont {Caprihan}},
  \bibinfo {author} {\bibfnamefont {S.~A.}\ \bibnamefont {Altobelli}}, \ and\
  \bibinfo {author} {\bibfnamefont {E.}~\bibnamefont {Fukushima}},\ }\href@noop
  {} {\bibfield  {journal} {\bibinfo  {journal} {Physical Review Letters}\
  }\textbf {\bibinfo {volume} {84}},\ \bibinfo {pages} {266} (\bibinfo {year}
  {2000})}\BibitemShut {NoStop}%
\bibitem [{\citenamefont {Mueth}\ \emph {et~al.}(2000)\citenamefont {Mueth},
  \citenamefont {Debregeas}, \citenamefont {Karczmar}, \citenamefont {Eng},
  \citenamefont {Nagel},\ and\ \citenamefont {Jaeger}}]{Mueth2000}%
  \BibitemOpen
  \bibfield  {author} {\bibinfo {author} {\bibfnamefont {D.~M.}\ \bibnamefont
  {Mueth}}, \bibinfo {author} {\bibfnamefont {G.~F.}\ \bibnamefont
  {Debregeas}}, \bibinfo {author} {\bibfnamefont {G.~S.}\ \bibnamefont
  {Karczmar}}, \bibinfo {author} {\bibfnamefont {P.~J.}\ \bibnamefont {Eng}},
  \bibinfo {author} {\bibfnamefont {S.~R.}\ \bibnamefont {Nagel}}, \ and\
  \bibinfo {author} {\bibfnamefont {H.~M.}\ \bibnamefont {Jaeger}},\
  }\href@noop {} {\bibfield  {journal} {\bibinfo  {journal} {Nature}\ }\textbf
  {\bibinfo {volume} {406}},\ \bibinfo {pages} {385} (\bibinfo {year}
  {2000})}\BibitemShut {NoStop}%
\bibitem [{\citenamefont {Seidler}\ \emph {et~al.}(2000)\citenamefont
  {Seidler}, \citenamefont {Martinez}, \citenamefont {Seeley}, \citenamefont
  {Kim}, \citenamefont {Behne}, \citenamefont {Zaranek}, \citenamefont
  {Chapman}, \citenamefont {Heald},\ and\ \citenamefont {Brewe}}]{Seidler2000}%
  \BibitemOpen
  \bibfield  {author} {\bibinfo {author} {\bibfnamefont {G.~T.}\ \bibnamefont
  {Seidler}}, \bibinfo {author} {\bibfnamefont {G.}~\bibnamefont {Martinez}},
  \bibinfo {author} {\bibfnamefont {L.~H.}\ \bibnamefont {Seeley}}, \bibinfo
  {author} {\bibfnamefont {K.~H.}\ \bibnamefont {Kim}}, \bibinfo {author}
  {\bibfnamefont {E.~A.}\ \bibnamefont {Behne}}, \bibinfo {author}
  {\bibfnamefont {S.}~\bibnamefont {Zaranek}}, \bibinfo {author} {\bibfnamefont
  {B.~D.}\ \bibnamefont {Chapman}}, \bibinfo {author} {\bibfnamefont {S.~M.}\
  \bibnamefont {Heald}}, \ and\ \bibinfo {author} {\bibfnamefont {D.~L.}\
  \bibnamefont {Brewe}},\ }\href@noop {} {\bibfield  {journal} {\bibinfo
  {journal} {Physical Review E}\ }\textbf {\bibinfo {volume} {62}},\ \bibinfo
  {pages} {8175} (\bibinfo {year} {2000})}\BibitemShut {NoStop}%
\bibitem [{\citenamefont {M\"{o}bius}\ \emph {et~al.}(2004)\citenamefont
  {M\"{o}bius}, \citenamefont {Cheng}, \citenamefont {Karczmar}, \citenamefont
  {Nagel},\ and\ \citenamefont {Jaeger}}]{Mobius2004}%
  \BibitemOpen
  \bibfield  {author} {\bibinfo {author} {\bibfnamefont {M.~E.}\ \bibnamefont
  {M\"{o}bius}}, \bibinfo {author} {\bibfnamefont {X.}~\bibnamefont {Cheng}},
  \bibinfo {author} {\bibfnamefont {G.~S.}\ \bibnamefont {Karczmar}}, \bibinfo
  {author} {\bibfnamefont {S.~R.}\ \bibnamefont {Nagel}}, \ and\ \bibinfo
  {author} {\bibfnamefont {H.~M.}\ \bibnamefont {Jaeger}},\ }\href@noop {}
  {\bibfield  {journal} {\bibinfo  {journal} {Physical Review Letters}\
  }\textbf {\bibinfo {volume} {93}},\ \bibinfo {pages} {198001} (\bibinfo
  {year} {2004})}\BibitemShut {NoStop}%
\bibitem [{\citenamefont {Aste}(2006)}]{Aste2006}%
  \BibitemOpen
  \bibfield  {author} {\bibinfo {author} {\bibfnamefont {T.}~\bibnamefont
  {Aste}},\ }\href@noop {} {\bibfield  {journal} {\bibinfo  {journal} {Physical
  Review Letters}\ }\textbf {\bibinfo {volume} {96}},\ \bibinfo {pages} {18002}
  (\bibinfo {year} {2006})}\BibitemShut {NoStop}%
\bibitem [{\citenamefont {Scheel}\ \emph {et~al.}(2008)\citenamefont {Scheel},
  \citenamefont {Seemann}, \citenamefont {Brinkmann}, \citenamefont
  {Di~Michiel}, \citenamefont {Sheppard},\ and\ \citenamefont
  {Herminghaus}}]{Scheel2008}%
  \BibitemOpen
  \bibfield  {author} {\bibinfo {author} {\bibfnamefont {M.}~\bibnamefont
  {Scheel}}, \bibinfo {author} {\bibfnamefont {R.}~\bibnamefont {Seemann}},
  \bibinfo {author} {\bibfnamefont {M.}~\bibnamefont {Brinkmann}}, \bibinfo
  {author} {\bibfnamefont {M.}~\bibnamefont {Di~Michiel}}, \bibinfo {author}
  {\bibfnamefont {A.}~\bibnamefont {Sheppard}}, \ and\ \bibinfo {author}
  {\bibfnamefont {S.}~\bibnamefont {Herminghaus}},\ }\href@noop {} {\bibfield
  {journal} {\bibinfo  {journal} {Journal of Physics-Condensed Matter}\
  }\textbf {\bibinfo {volume} {20}},\ \bibinfo {pages} {7} (\bibinfo {year}
  {2008})}\BibitemShut {NoStop}%
\bibitem [{\citenamefont {Jerkins}\ \emph {et~al.}(2008)\citenamefont
  {Jerkins}, \citenamefont {Schr\"oter}, \citenamefont {Swinney}, \citenamefont
  {Senden}, \citenamefont {Saadatfar},\ and\ \citenamefont
  {Aste}}]{Jerkins2008}%
  \BibitemOpen
  \bibfield  {author} {\bibinfo {author} {\bibfnamefont {M.}~\bibnamefont
  {Jerkins}}, \bibinfo {author} {\bibfnamefont {M.}~\bibnamefont {Schr\"oter}},
  \bibinfo {author} {\bibfnamefont {H.}~\bibnamefont {Swinney}}, \bibinfo
  {author} {\bibfnamefont {T.}~\bibnamefont {Senden}}, \bibinfo {author}
  {\bibfnamefont {M.}~\bibnamefont {Saadatfar}}, \ and\ \bibinfo {author}
  {\bibfnamefont {T.}~\bibnamefont {Aste}},\ }\href@noop {} {\bibfield
  {journal} {\bibinfo  {journal} {Physical Review Letters}\ }\textbf {\bibinfo
  {volume} {101}},\ \bibinfo {pages} {018301} (\bibinfo {year}
  {2008})}\BibitemShut {NoStop}%
\bibitem [{\citenamefont {Sanfratello}, \citenamefont {Fukushima},\ and\
  \citenamefont {Behringer}(2009)}]{Sanfratello2009}%
  \BibitemOpen
  \bibfield  {author} {\bibinfo {author} {\bibfnamefont {L.}~\bibnamefont
  {Sanfratello}}, \bibinfo {author} {\bibfnamefont {E.}~\bibnamefont
  {Fukushima}}, \ and\ \bibinfo {author} {\bibfnamefont {R.~P.}\ \bibnamefont
  {Behringer}},\ }\href@noop {} {\bibfield  {journal} {\bibinfo  {journal}
  {Granular Matter}\ }\textbf {\bibinfo {volume} {11}},\ \bibinfo {pages} {1}
  (\bibinfo {year} {2009})}\BibitemShut {NoStop}%
\bibitem [{\citenamefont {Zou}\ \emph {et~al.}(2009)\citenamefont {Zou},
  \citenamefont {Cheng}, \citenamefont {Rivers}, \citenamefont {Jaeger},\ and\
  \citenamefont {Nagel}}]{Zou2009}%
  \BibitemOpen
  \bibfield  {author} {\bibinfo {author} {\bibfnamefont {L.~N.}\ \bibnamefont
  {Zou}}, \bibinfo {author} {\bibfnamefont {X.}~\bibnamefont {Cheng}}, \bibinfo
  {author} {\bibfnamefont {M.~L.}\ \bibnamefont {Rivers}}, \bibinfo {author}
  {\bibfnamefont {H.~M.}\ \bibnamefont {Jaeger}}, \ and\ \bibinfo {author}
  {\bibfnamefont {S.~R.}\ \bibnamefont {Nagel}},\ }\href@noop {} {\bibfield
  {journal} {\bibinfo  {journal} {Science}\ }\textbf {\bibinfo {volume}
  {326}},\ \bibinfo {pages} {408} (\bibinfo {year} {2009})}\BibitemShut
  {NoStop}%
\bibitem [{\citenamefont {Jaoshvili}\ \emph {et~al.}(2010)\citenamefont
  {Jaoshvili}, \citenamefont {Esakia}, \citenamefont {Porrati},\ and\
  \citenamefont {Chaikin}}]{Jaoshvili2010}%
  \BibitemOpen
  \bibfield  {author} {\bibinfo {author} {\bibfnamefont {A.}~\bibnamefont
  {Jaoshvili}}, \bibinfo {author} {\bibfnamefont {A.}~\bibnamefont {Esakia}},
  \bibinfo {author} {\bibfnamefont {M.}~\bibnamefont {Porrati}}, \ and\
  \bibinfo {author} {\bibfnamefont {P.~M.}\ \bibnamefont {Chaikin}},\
  }\href@noop {} {\bibfield  {journal} {\bibinfo  {journal} {Physical Review
  Letters}\ }\textbf {\bibinfo {volume} {104}},\ \bibinfo {pages} {185501}
  (\bibinfo {year} {2010})}\BibitemShut {NoStop}%
\bibitem [{\citenamefont {Delaney}, \citenamefont {Di~Matteo},\ and\
  \citenamefont {Aste}(2010)}]{Delaney2010}%
  \BibitemOpen
  \bibfield  {author} {\bibinfo {author} {\bibfnamefont {G.~W.}\ \bibnamefont
  {Delaney}}, \bibinfo {author} {\bibfnamefont {T.}~\bibnamefont {Di~Matteo}},
  \ and\ \bibinfo {author} {\bibfnamefont {T.}~\bibnamefont {Aste}},\
  }\href@noop {} {\bibfield  {journal} {\bibinfo  {journal} {Soft Matter}\
  }\textbf {\bibinfo {volume} {6}},\ \bibinfo {pages} {2992} (\bibinfo {year}
  {2010})}\BibitemShut {NoStop}%
\bibitem [{\citenamefont {Higo}\ \emph {et~al.}(2011)\citenamefont {Higo},
  \citenamefont {Oka}, \citenamefont {Kimoto}, \citenamefont {Sanagawa},\ and\
  \citenamefont {Matsuhima}}]{Higo2011}%
  \BibitemOpen
  \bibfield  {author} {\bibinfo {author} {\bibfnamefont {Y.}~\bibnamefont
  {Higo}}, \bibinfo {author} {\bibfnamefont {F.}~\bibnamefont {Oka}}, \bibinfo
  {author} {\bibfnamefont {S.}~\bibnamefont {Kimoto}}, \bibinfo {author}
  {\bibfnamefont {T.}~\bibnamefont {Sanagawa}}, \ and\ \bibinfo {author}
  {\bibfnamefont {Y.}~\bibnamefont {Matsuhima}},\ }in\ \href@noop {} {\emph
  {\bibinfo {booktitle} {Advances in Bifurcation and Degradation in
  Geomaterials}}}\ (\bibinfo  {publisher} {Springer},\ \bibinfo {year} {2011})\
  pp.\ \bibinfo {pages} {37--43}\BibitemShut {NoStop}%
\bibitem [{\citenamefont {Shepherd}\ \emph {et~al.}(2012)\citenamefont
  {Shepherd}, \citenamefont {Conrad}, \citenamefont {Sabuwala}, \citenamefont
  {Gioia},\ and\ \citenamefont {Lewis}}]{Shepherd2012}%
  \BibitemOpen
  \bibfield  {author} {\bibinfo {author} {\bibfnamefont {R.~F.}\ \bibnamefont
  {Shepherd}}, \bibinfo {author} {\bibfnamefont {J.~C.}\ \bibnamefont
  {Conrad}}, \bibinfo {author} {\bibfnamefont {T.}~\bibnamefont {Sabuwala}},
  \bibinfo {author} {\bibfnamefont {G.~G.}\ \bibnamefont {Gioia}}, \ and\
  \bibinfo {author} {\bibfnamefont {J.~A.}\ \bibnamefont {Lewis}},\ }\href@noop
  {} {\bibfield  {journal} {\bibinfo  {journal} {Soft Matter}\ }\textbf
  {\bibinfo {volume} {8}},\ \bibinfo {pages} {4795} (\bibinfo {year}
  {2012})}\BibitemShut {NoStop}%
\bibitem [{\citenamefont {Fu}\ \emph {et~al.}(2012)\citenamefont {Fu},
  \citenamefont {Xi}, \citenamefont {Cao},\ and\ \citenamefont
  {Wang}}]{Fu2012}%
  \BibitemOpen
  \bibfield  {author} {\bibinfo {author} {\bibfnamefont {Y.}~\bibnamefont
  {Fu}}, \bibinfo {author} {\bibfnamefont {Y.}~\bibnamefont {Xi}}, \bibinfo
  {author} {\bibfnamefont {Y.~X.}\ \bibnamefont {Cao}}, \ and\ \bibinfo
  {author} {\bibfnamefont {Y.~J.}\ \bibnamefont {Wang}},\ }\href@noop {}
  {\bibfield  {journal} {\bibinfo  {journal} {Physical Review E}\ }\textbf
  {\bibinfo {volume} {85}},\ \bibinfo {pages} {051311} (\bibinfo {year}
  {2012})}\BibitemShut {NoStop}%
\bibitem [{\citenamefont {Brown}\ \emph {et~al.}(2012)\citenamefont {Brown},
  \citenamefont {Nasto}, \citenamefont {Athanassiadis},\ and\ \citenamefont
  {Jaeger}}]{Brown2012}%
  \BibitemOpen
  \bibfield  {author} {\bibinfo {author} {\bibfnamefont {E.}~\bibnamefont
  {Brown}}, \bibinfo {author} {\bibfnamefont {A.}~\bibnamefont {Nasto}},
  \bibinfo {author} {\bibfnamefont {A.~G.}\ \bibnamefont {Athanassiadis}}, \
  and\ \bibinfo {author} {\bibfnamefont {H.~M.}\ \bibnamefont {Jaeger}},\
  }\href@noop {} {\bibfield  {journal} {\bibinfo  {journal} {Phys Rev Lett}\
  }\textbf {\bibinfo {volume} {108}},\ \bibinfo {pages} {108302} (\bibinfo
  {year} {2012})}\BibitemShut {NoStop}%
\bibitem [{\citenamefont {Neudecker}\ \emph {et~al.}(2013)\citenamefont
  {Neudecker}, \citenamefont {Ulrich}, \citenamefont {Herminghaus},\ and\
  \citenamefont {Schr\"oter}}]{Neudecker2013}%
  \BibitemOpen
  \bibfield  {author} {\bibinfo {author} {\bibfnamefont {M.}~\bibnamefont
  {Neudecker}}, \bibinfo {author} {\bibfnamefont {S.}~\bibnamefont {Ulrich}},
  \bibinfo {author} {\bibfnamefont {S.}~\bibnamefont {Herminghaus}}, \ and\
  \bibinfo {author} {\bibfnamefont {M.}~\bibnamefont {Schr\"oter}},\ }\href
  {\doibase 10.1103/PhysRevLett.111.028001} {\bibfield  {journal} {\bibinfo
  {journal} {Phys. Rev. Lett.}\ }\textbf {\bibinfo {volume} {111}},\ \bibinfo
  {pages} {028001} (\bibinfo {year} {2013})}\BibitemShut {NoStop}%
\bibitem [{\citenamefont {Baxter}\ \emph {et~al.}(1989)\citenamefont {Baxter},
  \citenamefont {Behringer}, \citenamefont {Fagert},\ and\ \citenamefont
  {Johnson}}]{Baxter1989}%
  \BibitemOpen
  \bibfield  {author} {\bibinfo {author} {\bibfnamefont {G.~W.}\ \bibnamefont
  {Baxter}}, \bibinfo {author} {\bibfnamefont {R.~P.}\ \bibnamefont
  {Behringer}}, \bibinfo {author} {\bibfnamefont {T.}~\bibnamefont {Fagert}}, \
  and\ \bibinfo {author} {\bibfnamefont {G.~A.}\ \bibnamefont {Johnson}},\
  }\href@noop {} {\bibfield  {journal} {\bibinfo  {journal} {Physical Review
  Letters}\ }\textbf {\bibinfo {volume} {62}},\ \bibinfo {pages} {2825}
  (\bibinfo {year} {1989})}\BibitemShut {NoStop}%
\bibitem [{\citenamefont {Royer}\ \emph {et~al.}(2005)\citenamefont {Royer},
  \citenamefont {Corwin}, \citenamefont {Flior}, \citenamefont {Cordero},
  \citenamefont {Rivers}, \citenamefont {Eng},\ and\ \citenamefont
  {Jaeger}}]{Royer2005}%
  \BibitemOpen
  \bibfield  {author} {\bibinfo {author} {\bibfnamefont {J.~R.}\ \bibnamefont
  {Royer}}, \bibinfo {author} {\bibfnamefont {E.~I.}\ \bibnamefont {Corwin}},
  \bibinfo {author} {\bibfnamefont {A.}~\bibnamefont {Flior}}, \bibinfo
  {author} {\bibfnamefont {M.~L.}\ \bibnamefont {Cordero}}, \bibinfo {author}
  {\bibfnamefont {M.~L.}\ \bibnamefont {Rivers}}, \bibinfo {author}
  {\bibfnamefont {P.~J.}\ \bibnamefont {Eng}}, \ and\ \bibinfo {author}
  {\bibfnamefont {H.~M.}\ \bibnamefont {Jaeger}},\ }\href@noop {} {\bibfield
  {journal} {\bibinfo  {journal} {Nature Physics}\ }\textbf {\bibinfo {volume}
  {1}},\ \bibinfo {pages} {164} (\bibinfo {year} {2005})}\BibitemShut {NoStop}%
\bibitem [{\citenamefont {Royer}\ \emph {et~al.}(2007)\citenamefont {Royer},
  \citenamefont {Corwin}, \citenamefont {Eng},\ and\ \citenamefont
  {Jaeger}}]{Royer2007}%
  \BibitemOpen
  \bibfield  {author} {\bibinfo {author} {\bibfnamefont {J.~R.}\ \bibnamefont
  {Royer}}, \bibinfo {author} {\bibfnamefont {E.~I.}\ \bibnamefont {Corwin}},
  \bibinfo {author} {\bibfnamefont {P.~J.}\ \bibnamefont {Eng}}, \ and\
  \bibinfo {author} {\bibfnamefont {H.~M.}\ \bibnamefont {Jaeger}},\
  }\href@noop {} {\bibfield  {journal} {\bibinfo  {journal} {Physical Review
  Letters}\ }\textbf {\bibinfo {volume} {99}},\ \bibinfo {pages} {038003}
  (\bibinfo {year} {2007})}\BibitemShut {NoStop}%
\bibitem [{\citenamefont {Maladen}\ \emph {et~al.}(2009)\citenamefont
  {Maladen}, \citenamefont {Ding}, \citenamefont {Li},\ and\ \citenamefont
  {Goldman}}]{Maladen2009}%
  \BibitemOpen
  \bibfield  {author} {\bibinfo {author} {\bibfnamefont {R.~D.}\ \bibnamefont
  {Maladen}}, \bibinfo {author} {\bibfnamefont {Y.}~\bibnamefont {Ding}},
  \bibinfo {author} {\bibfnamefont {C.}~\bibnamefont {Li}}, \ and\ \bibinfo
  {author} {\bibfnamefont {D.~I.}\ \bibnamefont {Goldman}},\ }\href@noop {}
  {\bibfield  {journal} {\bibinfo  {journal} {Science}\ }\textbf {\bibinfo
  {volume} {325}},\ \bibinfo {pages} {314} (\bibinfo {year}
  {2009})}\BibitemShut {NoStop}%
\bibitem [{\citenamefont {Royer}\ \emph {et~al.}(2011)\citenamefont {Royer},
  \citenamefont {Conyers}, \citenamefont {Corwin}, \citenamefont {Eng},\ and\
  \citenamefont {Jaeger}}]{Royer2011}%
  \BibitemOpen
  \bibfield  {author} {\bibinfo {author} {\bibfnamefont {J.~R.}\ \bibnamefont
  {Royer}}, \bibinfo {author} {\bibfnamefont {B.}~\bibnamefont {Conyers}},
  \bibinfo {author} {\bibfnamefont {E.~I.}\ \bibnamefont {Corwin}}, \bibinfo
  {author} {\bibfnamefont {P.~J.}\ \bibnamefont {Eng}}, \ and\ \bibinfo
  {author} {\bibfnamefont {H.~M.}\ \bibnamefont {Jaeger}},\ }\href@noop {}
  {\bibfield  {journal} {\bibinfo  {journal} {Europhysics Letters}\ }\textbf
  {\bibinfo {volume} {93}},\ \bibinfo {pages} {28008} (\bibinfo {year}
  {2011})}\BibitemShut {NoStop}%
\bibitem [{\citenamefont {Waitukaitis}\ and\ \citenamefont
  {Jaeger}(2012)}]{Waitukaitis2012}%
  \BibitemOpen
  \bibfield  {author} {\bibinfo {author} {\bibfnamefont {S.~R.}\ \bibnamefont
  {Waitukaitis}}\ and\ \bibinfo {author} {\bibfnamefont {H.~M.}\ \bibnamefont
  {Jaeger}},\ }\href@noop {} {\bibfield  {journal} {\bibinfo  {journal}
  {Nature}\ }\textbf {\bibinfo {volume} {487}},\ \bibinfo {pages} {205}
  (\bibinfo {year} {2012})}\BibitemShut {NoStop}%
\bibitem [{\citenamefont {Fahrig}\ \emph {et~al.}(1997)\citenamefont {Fahrig},
  \citenamefont {Fox}, \citenamefont {Lownie},\ and\ \citenamefont
  {Holdsworth}}]{Fahrig1997}%
  \BibitemOpen
  \bibfield  {author} {\bibinfo {author} {\bibfnamefont {R.}~\bibnamefont
  {Fahrig}}, \bibinfo {author} {\bibfnamefont {A.~J.}\ \bibnamefont {Fox}},
  \bibinfo {author} {\bibfnamefont {S.}~\bibnamefont {Lownie}}, \ and\ \bibinfo
  {author} {\bibfnamefont {D.~W.}\ \bibnamefont {Holdsworth}},\ }\href@noop {}
  {\bibfield  {journal} {\bibinfo  {journal} {American Journal of
  Neuroradiology}\ }\textbf {\bibinfo {volume} {18}},\ \bibinfo {pages} {1507}
  (\bibinfo {year} {1997})}\BibitemShut {NoStop}%
\bibitem [{\citenamefont {Orth}, \citenamefont {Wallace},\ and\ \citenamefont
  {Kuo}(2008)}]{Orth2008}%
  \BibitemOpen
  \bibfield  {author} {\bibinfo {author} {\bibfnamefont {R.~C.}\ \bibnamefont
  {Orth}}, \bibinfo {author} {\bibfnamefont {M.~J.}\ \bibnamefont {Wallace}}, \
  and\ \bibinfo {author} {\bibfnamefont {M.~D.}\ \bibnamefont {Kuo}},\
  }\href@noop {} {\bibfield  {journal} {\bibinfo  {journal} {Journal of
  Vascular and Interventional Radiology}\ }\textbf {\bibinfo {volume} {19}},\
  \bibinfo {pages} {814} (\bibinfo {year} {2008})}\BibitemShut {NoStop}%
\bibitem [{DCM()}]{DCMTK}%
  \BibitemOpen
  \href@noop {} {}\bibinfo {note} {DICOM Toolkit (DCMTK).
  \url{http://dicom.offis.de/dcmtk.php.en} (Accessed Aug. 8,
  2013).}\BibitemShut {Stop}%
\bibitem [{\citenamefont {Yang}\ \emph {et~al.}(2006)\citenamefont {Yang},
  \citenamefont {Kwan}, \citenamefont {Miller},\ and\ \citenamefont
  {Boone}}]{Yang2006}%
  \BibitemOpen
  \bibfield  {author} {\bibinfo {author} {\bibfnamefont {K.}~\bibnamefont
  {Yang}}, \bibinfo {author} {\bibfnamefont {A.~L.~C.}\ \bibnamefont {Kwan}},
  \bibinfo {author} {\bibfnamefont {D.~F.}\ \bibnamefont {Miller}}, \ and\
  \bibinfo {author} {\bibfnamefont {J.~M.}\ \bibnamefont {Boone}},\ }\href@noop
  {} {\bibfield  {journal} {\bibinfo  {journal} {Medical Physics}\ }\textbf
  {\bibinfo {volume} {33}},\ \bibinfo {pages} {1695} (\bibinfo {year}
  {2006})}\BibitemShut {NoStop}%
\bibitem [{\citenamefont {Feldkamp}, \citenamefont {Davis},\ and\ \citenamefont
  {Kress}(1984)}]{FDKOrig}%
  \BibitemOpen
  \bibfield  {author} {\bibinfo {author} {\bibfnamefont {L.~A.}\ \bibnamefont
  {Feldkamp}}, \bibinfo {author} {\bibfnamefont {L.}~\bibnamefont {Davis}}, \
  and\ \bibinfo {author} {\bibfnamefont {J.}~\bibnamefont {Kress}},\
  }\href@noop {} {\bibfield  {journal} {\bibinfo  {journal} {J. Opt. Soc. Am.
  A}\ }\textbf {\bibinfo {volume} {1}},\ \bibinfo {pages} {612} (\bibinfo
  {year} {1984})}\BibitemShut {NoStop}%
\bibitem [{\citenamefont {Criminisi}, \citenamefont {Perez},\ and\
  \citenamefont {Toyama}(2004)}]{Criminisi:Inpainting}%
  \BibitemOpen
  \bibfield  {author} {\bibinfo {author} {\bibfnamefont {A.}~\bibnamefont
  {Criminisi}}, \bibinfo {author} {\bibfnamefont {P.}~\bibnamefont {Perez}}, \
  and\ \bibinfo {author} {\bibfnamefont {K.}~\bibnamefont {Toyama}},\
  }\href@noop {} {\bibfield  {journal} {\bibinfo  {journal} {IEEE Trans Image
  Process}\ }\textbf {\bibinfo {volume} {13}},\ \bibinfo {pages} {1200}
  (\bibinfo {year} {2004})}\BibitemShut {NoStop}%
\bibitem [{\citenamefont {Bradski}(2000)}]{opencv_library}%
  \BibitemOpen
  \bibfield  {author} {\bibinfo {author} {\bibfnamefont {G.}~\bibnamefont
  {Bradski}},\ }\href@noop {} {\bibfield  {journal} {\bibinfo  {journal}
  {Doctor Dobbs Journal}\ }\textbf {\bibinfo {volume} {25}},\ \bibinfo {pages}
  {120} (\bibinfo {year} {2000})}\BibitemShut {NoStop}%
\bibitem [{\citenamefont {Kak}\ \emph {et~al.}(1988)\citenamefont {Kak},
  \citenamefont {Slaney}, \citenamefont {in~Medicine},\ and\ \citenamefont
  {Society}}]{KakSlaney}%
  \BibitemOpen
  \bibfield  {author} {\bibinfo {author} {\bibfnamefont {A.~C.}\ \bibnamefont
  {Kak}}, \bibinfo {author} {\bibfnamefont {M.}~\bibnamefont {Slaney}},
  \bibinfo {author} {\bibfnamefont {I.~E.}\ \bibnamefont {in~Medicine}}, \ and\
  \bibinfo {author} {\bibfnamefont {B.}~\bibnamefont {Society}},\ }\href@noop
  {} {\emph {\bibinfo {title} {Principles of computerized tomographic
  imaging}}}\ (\bibinfo  {publisher} {IEEE Press},\ \bibinfo {address} {New
  York},\ \bibinfo {year} {1988})\ \bibinfo {note}
  {\url{http://www.slaney.org/pct/pct-toc.html}}\BibitemShut {NoStop}%
\bibitem [{\citenamefont {Oliphant}(2007)}]{SciPyNumPy}%
  \BibitemOpen
  \bibfield  {author} {\bibinfo {author} {\bibfnamefont {T.~E.}\ \bibnamefont
  {Oliphant}},\ }\href {http://link.aip.org/link/?CSX/9/10/1} {\bibfield
  {journal} {\bibinfo  {journal} {Computing in Science \& Engineering}\
  }\textbf {\bibinfo {volume} {9}},\ \bibinfo {pages} {10} (\bibinfo {year}
  {2007})}\BibitemShut {NoStop}%
\bibitem [{\citenamefont {Hunter}(2007)}]{matplotlib}%
  \BibitemOpen
  \bibfield  {author} {\bibinfo {author} {\bibfnamefont {J.~D.}\ \bibnamefont
  {Hunter}},\ }\href {http://link.aip.org/link/?CSX/9/90/1} {\bibfield
  {journal} {\bibinfo  {journal} {Computing in Science \& Engineering}\
  }\textbf {\bibinfo {volume} {9}},\ \bibinfo {pages} {90} (\bibinfo {year}
  {2007})}\BibitemShut {NoStop}%
\bibitem [{\citenamefont {Coelho}(2013)}]{mahotas}%
  \BibitemOpen
  \bibfield  {author} {\bibinfo {author} {\bibfnamefont {L.~P.}\ \bibnamefont
  {Coelho}},\ }\href {\doibase http://dx.doi.org/10.5334/jors.ac} {\bibfield
  {journal} {\bibinfo  {journal} {Journal of Open Research Software}\ }\textbf
  {\bibinfo {volume} {1}} (\bibinfo {year} {2013}),\
  http://dx.doi.org/10.5334/jors.ac}\BibitemShut {NoStop}%
\bibitem [{\citenamefont {Prince}\ and\ \citenamefont
  {Links}(2006)}]{PrinceLinks}%
  \BibitemOpen
  \bibfield  {author} {\bibinfo {author} {\bibfnamefont {J.~L.}\ \bibnamefont
  {Prince}}\ and\ \bibinfo {author} {\bibfnamefont {J.~M.}\ \bibnamefont
  {Links}},\ }\href@noop {} {\emph {\bibinfo {title} {Medical Imaging: Signals
  and Systems}}},\ Pearson Prentice Hall Bioengineering\ (\bibinfo  {publisher}
  {Pearson Education},\ \bibinfo {address} {Upper Saddle River, NJ},\ \bibinfo
  {year} {2006})\BibitemShut {NoStop}%
\bibitem [{\citenamefont {Hsieh}(2009)}]{Hsieh2009}%
  \BibitemOpen
  \bibfield  {author} {\bibinfo {author} {\bibfnamefont {J.}~\bibnamefont
  {Hsieh}},\ }\href
  {http://libproxy.mit.edu/login?url=http://search.ebscohost.com/login.aspx?direct=true&db=cat00916a&AN=mit.001703962&site=eds-live}
  {\emph {\bibinfo {title} {Computed tomography [electronic resource] :
  principles, design, artifacts, and recent advances}}},\ SPIE Press monograph:
  PM188\ (\bibinfo  {publisher} {SPIE},\ \bibinfo {address} {Bellingham,
  Wash.},\ \bibinfo {year} {2009})\BibitemShut {NoStop}%
\bibitem [{\citenamefont {Reiser}(2009)}]{Reiser2009}%
  \BibitemOpen
  \bibfield  {author} {\bibinfo {author} {\bibfnamefont {M.}~\bibnamefont
  {Reiser}},\ }\href@noop {} {\emph {\bibinfo {title} {Multislice CT
  [electronic resource]}}},\ Medical radiology\ (\bibinfo  {publisher}
  {Springer},\ \bibinfo {address} {Berlin},\ \bibinfo {year}
  {2009})\BibitemShut {NoStop}%
\end{thebibliography}%

\end{document}